# Artificial Intelligence-Generated Terahertz Multi-Resonant Metasurfaces via Improved Transformer and CGAN Neural Networks

Yangpeng Huang, Naixing Feng, *Senior Member, IEEE*, and Yijun Cai

*Abstract*—It is well known that the inverse design of terahertz (THz) multi-resonant graphene metasurfaces by using traditional deep neural networks (DNNs) has limited generalization ability. In this paper, we propose improved Transformer and conditional generative adversarial neural networks (CGAN) for the inverse design of graphene metasurfaces based upon THz multi-resonant absorption spectra. The improved Transformer can obtain higher accuracy and generalization performance in the StoV (Spectrum to Vector) design compared to traditional multilayer perceptron (MLP) neural networks, while the StoI (Spectrum to Image) design achieved through CGAN can provide more comprehensive information and higher accuracy than the StoV design obtained by MLP. Moreover, the improved CGAN can achieve the inverse design of graphene metasurface images directly from the desired multi-resonant absorption spectra. It is turned out that this work can finish facilitating the design process of artificial intelligence-generated metasurfaces (AIGM), and even provide a useful guide for developing complex THz metasurfaces based on 2D materials using generative neural networks.

*Index Terms*— Conditional generative adversarial neural networks (CGAN), inverse design, improved Transformer, multi-resonant metasurface, terahertz.

## I. INTRODUCTION

Terahertz (THz) technology has garnered a great deal of research interest due to its unique technical characteristics [1]-[5]. As known to us, the graphene-based THz metasurfaces, with their distinctive optical properties, have hitherto enabled the development and improvement of numerous applications including optical detection [6], imaging [7], [8], sensing [9], and tunable absorption [10]. The design of metasurfaces often relies on the desired THz spectrum, which requires extensive simulation and continuous trial and error by experienced experts. However, due to the limitations of computational performance, only a limited number of design parameters could be allowed to adjust, making it difficult to obtain the optimal structure [11], [12]. For a long time, traditional algorithms like adjoint methods [13], topology optimization [14], and genetic algorithms [15] have been used to solve the inverse design problem in metasurfaces. To the best of our knowledge, nevertheless, these traditional methods typically require significant computing power and time. Furthermore, as the number of optimization parameters and spatial dimension increases, the required computing power also increases, which is not conducive to the rapid development of metasurfaces.

As we all know, deep learning (DL) is widely recognized for its unique advantage of being data-driven, enabling models to automatically extract valuable information from large amounts of data [16]-[19]. As an end-to-end inverse design approach, the DL demonstrates outstanding performance in designing metasurfaces. It not only eliminates the need for trial-and-error experiences in numerical simulations, but also effectively addresses the computational challenges faced by traditional methods. Furthermore, the use of DL algorithms can explore a large number of design parameters and obtain optimal structures that cannot be achieved by traditional algorithms. DL technology has been widely applied in the inverse design of various nanostructures, such as multilayer nanoparticles [20], multilayer thin films [21], metamaterials [22], metasurfaces [23]. Especially for the inverse design problems of graphene-based metasurfaces, the DL technology can exhibit greater advantages upon conventional algorithms considering both the efficiency and accuracy. Next, Harper et al. developed artificial neural networks (ANNs) that can relate metasurface geometries to reflection and transmission spectra, enabling the neural networks to suggest device geometries based on a desired optical performance. This inverts the design process for metasurfaces [24]. Lin *et al.* proposed an improved transfer function (TF)-based ANN model that can directly generate structure parameters to match the customer-expected THz transmission spectrum, using an inner ring resonator and a split outer ring resonator [25]. Du *et al.* proposed a scalable multi-task learning neural network that can capture the impact of nanostructure dimensions on their optical absorption in graphene-based metasurfaces, obtained through inverse design of absorption spectrum in the visible frequency [26]. Liu *et al*. constructed a generative network which can produce candidate patterns that match the desired spectrum with high fidelity. The generative network can produce a specific metasurface

This work was supported in part by the National Nature Science Foundation (NSF) of China under Grant 62005232, 62271001, by the Natural Science Foundation of Fujian Province under Grant 2020J01294, by the Anhui Provincial Natural Science Foundation under Grant 2022AH030014, and by the Graduate Science and Technology Innovation Projects under Grant YKJCX2022071. *(Corresponding author: Yijun Cai) (Equally contributed authors: Yangpeng Huang and Naixing Feng)*
Y. Cai and Y. Huang are with Smart Sensing Integrated Circuit Engineering Research Center of Universities in Fujian Province, Xiamen University of Technology, Xiamen 361000, China (e-mails: yijuncai@foxmail.com; huangyangpeng@s.xmut.edu.cn).
N. Feng is with the Key Laboratory of Intelligent Computing and Signal Processing, Ministry of Education, Anhui University, Hefei 230601, China, and Information Materials and Intelligent Sensing Laboratory of Anhui Province, Anhui University, Hefei 230601, China (e-mails: fengnaixing@gmail.com).

corresponding to a given input transmission spectrum in the THz range [27].

To the best of our knowledge, the inverse design approaches described in previous work have primarily used conventional neural networks. Furthermore, the inverse design of graphene-based metasurfaces primarily relies on structural parameters, which result in limited information content, generalization capacity, and practical limitations. As a result, there is an urgent need to develop the neural network model with higher generalization capacity and improved accuracy for the inverse design of THz multi-resonant metasurfaces based on graphene. In our previous work, we employed an improved Transformer neural network for achieving the inverse design of multilayer metamaterials, which can obtain significant improvements in both generalization ability and accuracy [28]. However, it was only applicable to the one-dimensional layered structures and powerless for more complex metasurface structures.

In this article, we propose the improved Transformer and conditional generative adversarial neural networks (CGAN) for the inverse design of graphene metasurfaces based on THz multi-resonant absorption spectra. The inputs for both neural networks are the THz multi-resonant absorption spectrum, and the outputs are the parameter vector and the pattern image of the graphene metasurface, referred to as StoV (Spectrum to Vector) and StoI (Spectrum to Image), respectively, in the remaining parts of the paper. Besides, as compared with the traditional multilayer perceptron (MLP) neural networks, the improved Transformer can achieve both higher accuracy and generalization performance in the StoV design. Furthermore, the StoI design achieved through CGAN can provide more comprehensive information and higher accuracy than the StoV design obtained by the traditional MLP neural network, and it truly facilitates the design process of AIGM (AI-Generated Metasurface).

## II. STRUCTURE AND PARAMETERS

To verify the effectiveness of the proposed inverse design algorithm, we constructed a graphene metasurface composed of graphene strips with different widths, as shown in Fig. 1. It consists of a monolayer graphene metasurface, SiC dielectric layer and conventional perfect electric conductor (PEC) substrate, where the thickness of the dielectric layer SiC $h$ is 2.8 $\mu$m and the periodic width $P$ is 20 $\mu$m. Graphene strips with different widths generate different resonances, making the device suitable for use as a multi-resonant THz resonator. To facilitate the subsequent design, we divided the 20 $\mu$m period width of the multi-resonant metasurface into 20 strips, each with a width of 1 $\mu$m. There are 6 different cases of graphene chemical potential, including 0 eV, 0.6 eV, 0.7 eV, 0.8 eV, 0.9 eV, and 1.0 eV, where 0 eV represents the absence of graphene on the surface of the strip.

In the design, the refractive index of SiC is set as 2.5 according to [29], [30] and the refractive index of graphene can be calculated using its surface conductivity, as defined in (1)-(6) [28], [31]:

$$\sigma(\omega, \mu_c, \Gamma, T) = \sigma_{\text{intra}} + \sigma_{\text{inter}}, \quad (1)$$

$$\sigma_{\text{intra}} = \frac{je^2}{\pi\hbar^2(\omega-j2\Gamma)} \int_0^\infty \xi \left( \frac{\partial f_d(\xi,\mu_c,T)}{\partial \xi} - \frac{\partial f_d(-\xi,\mu_c,T)}{\partial \xi} \right) d\xi, \quad (2)$$

$$\sigma_{\text{inter}} = -\frac{je^2(\omega-j2\Gamma)}{\pi\hbar^2} \int_0^\infty \frac{f_d(-\xi,\mu_c,T)-f_d(\xi,\mu_c,T)}{(\omega-j2\Gamma)^2-4(\xi/\hbar)^2} d\xi, \quad (3)$$

$$f_d(\xi, \mu_c, T) = \left(e^{(\xi-\mu_c)/k_BT} + 1\right)^{-1}, \quad (4)$$

$$\varepsilon_{in}(\omega) = \varepsilon_{xx}(\omega) = \varepsilon_{yy}(\omega) = \varepsilon_0 + i\frac{\sigma}{\Delta\omega}, \quad (5)$$

$$n_g = (\varepsilon_{in}\mu_r)^{1/2}. \quad (6)$$

Here, the symbols $\sigma_{intra}$ and $\sigma_{inter}$ denote the conductivity of graphene resulting from the intraband and interband transitions, respectively. $\hbar$, $\Gamma$, $\omega$, $e$, $k_B$, $T$, $\mu_c$ and $\xi$ denote the reduced Planck constant, the scattering rate, the radian frequency, the electron charge, the Boltzmann constant, the temperature of Kelvin, the chemical potential and the energy of electrons, respectively. The Fermi-Dirac distribution is represented by $f_d(\xi, \mu_c, T)$. The in-plane components of graphene relative permittivity are represented by $\varepsilon_{in}$, $\varepsilon_{xx}$ and $\varepsilon_{yy}$ and the permittivity of vacuum is represented by $\varepsilon_0$. $\Delta$ denotes the thickness of graphene. $n_g$ and $\mu_r$ are the refractive index and relative permeability of graphene, respectively. Moreover, $\mu_r$ is assumed as 1 due to its nonmagnetic property.

In order to obtain the absorption spectrum of THz multi-resonant metasurfaces based on graphene, we employed simulation using the numerical method of finite element method (FEM). In our simulation, we applied periodic boundary conditions in the *x*-axis and *y*-axis directions and imposed THz incidence downward from the top surface of the graphene metasurface. To capture the localized enhanced electromagnetic fields of the graphene layer, meshes of user-defined size were applied, while tetrahedral meshes were used for the remaining domains of the structure. The simulation was performed on a metasurface with a period width of 20 $\mu$m, which was divided into 20 strips, each with six different values of chemical potential. Besides, we employed uniform sampling to generate 20,000 data sets, with 19,000 used for training and 1,000 for testing to avoid solution space bias. The data sets consisted of different combinations of graphene chemical potentials and their corresponding absorption spectra.

## III. StoV DESIGN FOR GRAPHENE METASURFACE

We first consider the StoV inverse design for the THz multi-resonant metasurface with our proposed improved Transformer network. The graphene metasurface with a periodic 20 $\mu$m is divided into 20 strips, and the chemical potential parameters of the 20 graphene strips are represented using the vector $C = [c_1, c_2, ..., c_{20}]$, where $c_1, c_2, ...c_{20}$ represents the chemical potential of each graphene strip.

Electrical biasing or chemical doping can be used to tune the chemical potential and design the diversity of THz multi-resonant metasurfaces based on graphene. The green and red curve samples in Fig. 2 correspond to Segments I and II and have chemical potential representations of $C_I$=[0, 0.9, 0, 0.9, 0.9, 0, 0.9, 0.9, 0, 0, 0, 0, 0, 0, 0, 0, 0, 0, 0, 0] eV and $C_{II}$=[0, 0.7, 0, 1.0, 1.0, 0, 0, 1.0, 1.0, 1.0, 0, 0, 0.9, 0.9, 0.9, 0.9, 0.9, 0.9, 0, 0, 0] eV, respectively. As shown in Fig. 2, these two samples have two and four resonant absorption peaks, respectively.

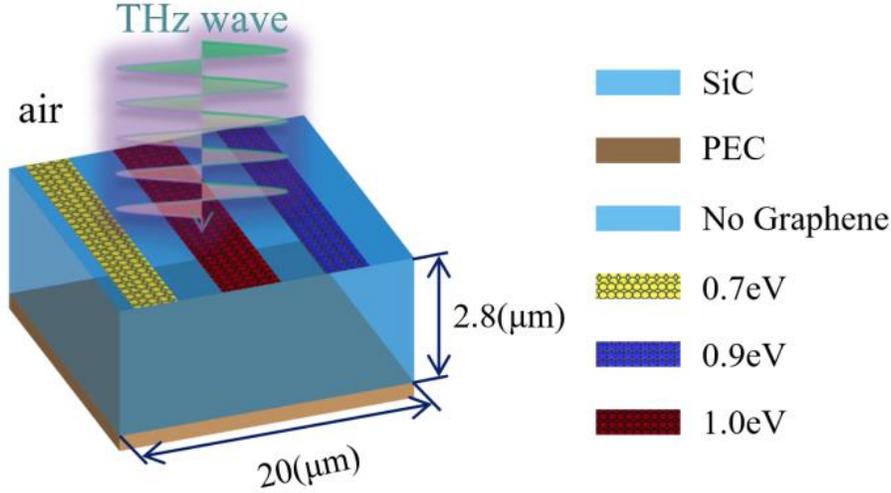

Fig. 1. Schematic diagram of the proposed THz multi-resonant graphene metasurface.

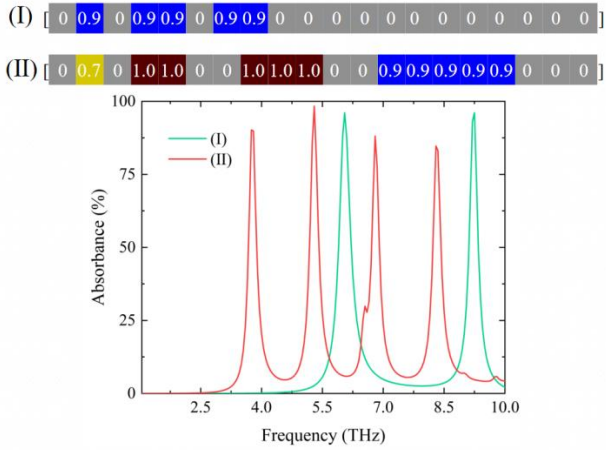

Fig. 2. Two data set examples for StoV design: two segments of graphene metasurface and their corresponding multi-resonant absorption spectra.

Due to the fact that some simulated absorption spectra do not perform well, and real-world fabrication conditions must be taken into account, such as ensuring that the chemical potentials of adjacent graphene strips are the same, collecting data sets for THz multi-resonant metasurfaces based on graphene is challenging. This can result in a smaller amount of data than expected, which can easily lead to overfitting in larger neural networks. To solve this problem, we propose an improved Transformer network to design the desired THz absorption spectrum based on the training data $D_i = \{(S_i, C_j), i = 1,2…n, j = 1,2…h\}$, where $S_i$ and $C_j$ represent the sampling value of THz multi-resonant absorption spectrum and the chemical potential of 20 graphene strips in the metasurface, respectively. In the proposed improved Transformer model, the THz multi-resonant absorption spectrum is used as the input to the neural network, which consists of 181 sampling points representing frequencies ranging from 1 THz to 10 THz with an interval of 0.05 THz. The output of the network is the chemical potential vector of the 20 strips in the graphene metasurface.

As seen in Fig. 3, like the original Transformer [32], our proposed improved Transformer network employs the classical Encoder-Decoder model. The Encoder and Decoder have identical structures. In the Encoder module, each input vector is linearly mapped into three vectors $Q$, $K$, $V$. $Q$ and $K$ are used to compute the attention map of the Encoder. Encoder is performed for $[a^1, a^2, …, a^{181}]$ by means of adopting the self-attention mechanism. Firstly, $Q$, $K$ and $V$ matrices are computed by three FCLs, see (7)-(10), where $W^q$, $W^k$, $W^v$ are the weights in three FCLs, respectively. The $[q^1, q^2, …, q^{181}]$, $[k^1, k^2, …, k^{181}]$, $[v^1, v^2, …, v^{181}]$ are the results of the linear mapping of $[a^1, a^2, …, a^{181}]$ in three FCLs, respectively.

$$q^i = W^q a^i, k^i = W^k a^i, v^i = W^v a^i, \quad (7)$$

$$Q = [q^1 q^2 … q^{181}] = W^q [a^1 a^2 … a^{181}], \quad (8)$$

$$K = [k^1 k^2 … k^{181}] = W^k [a^1 a^2 … a^{181}], \quad (9)$$

$$V = [v^1 v^2 … v^{181}] = W^v [a^1 a^2 … a^{181}]. \quad (10)$$

Then we compute the matrix of self-attention outputs through $Q$, $K$ and $V$, as illustrated in (11)-(13). $A$ is the attention score matrix, while $B$ is the matrix of the self-attention output.

$$A = K^T Q, \ B = VA, \quad (11)$$

$$A = \begin{bmatrix} a_{(1,1)} & \cdots & a_{(1,181)} \\ \vdots & \ddots & \vdots \\ a_{(181,1)} & \cdots & a_{(181,181)} \end{bmatrix} = \begin{bmatrix} k^1 \\ \vdots \\ k^{181} \end{bmatrix} [q^1 … q^{181}], \quad (12)$$

$$B = \begin{bmatrix} b_{(1,1)} & \cdots & b_{(1,181)} \\ \vdots & \ddots & \vdots \\ b_{(181,1)} & \cdots & b_{(181,181)} \end{bmatrix} = [v^1 … v^{181}] \begin{bmatrix} a_{(1,1)} & \cdots & a_{(1,181)} \\ \vdots & \ddots & \vdots \\ a_{(181,1)} & \cdots & a_{(181,181)} \end{bmatrix}. \quad (13)$$

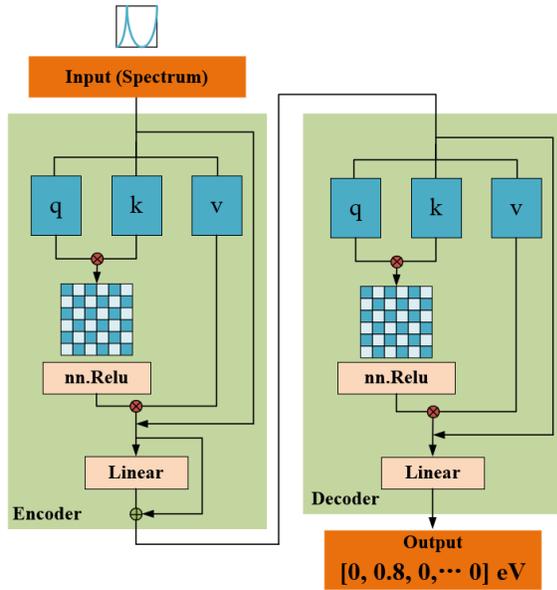

Fig. 3. Structure of the proposed improved Transformer Network.

To train and validate the improved Transformer in our work, we choose the Adam optimizer, while the batchsize is set as 256, the learning rate is set as 0.001 and the weight decay is set as 1e-5. For the testing set, we define a function to calculate the prediction accuracy of neural networks as shown in (14):

$$H = \frac{\sqrt{\sum_{i=1}^{n}(p_i-x_i)^2}}{\sqrt{\sum_{i=1}^{n}p_i^2}}. \quad (14)$$

where $H$, $p_i$, $x_i$ and $n$ denote the average relative accuracy, the sampling point of the target absorption spectrum, the sampling value of the predicted absorption spectrum and the total amount of sampling points, respectively. In this work, Equation (14) is utilized to evaluate the test results of the improved Transformer and the traditional MLP networks.

It is obvious that the inverse design accuracy based on the improved Transformer is higher than that based on the traditional MLP for the design of StoV. As reflected in Fig. 4(a), the Transformer converges at epoch=16, while the MLP converges at epoch=33. Moreover, the training and testing loss after Transformer converges is 1.13, while the training and testing loss after MLP converges is 1.87. Next, in Fig. 4(b), the test accuracy of Transformer reaches 96.14% after convergence, while that of MLP is only 94.27%. Therefore, the proposed network can exhibit higher accuracy and faster convergence speed than the traditional neural network.

To further confirm the superiority of the improved Transformer in the application of inverse design, we compare the target spectrum with the predicted spectrum predicted by means of the standard MLP and the improved Transformer networks, seen in Fig. 5. Obviously, due to the introduction of self-attention mechanism, Transformer is able to better extract the key information from the data, capture the internal relevance of the data or features, ignore the unimportant information, and reduce the requirement of the dataset. Transformer neural network is able to achieve higher accuracy and better generalization capability in the inverse design of THz multi-resonant metasurface based on graphene.

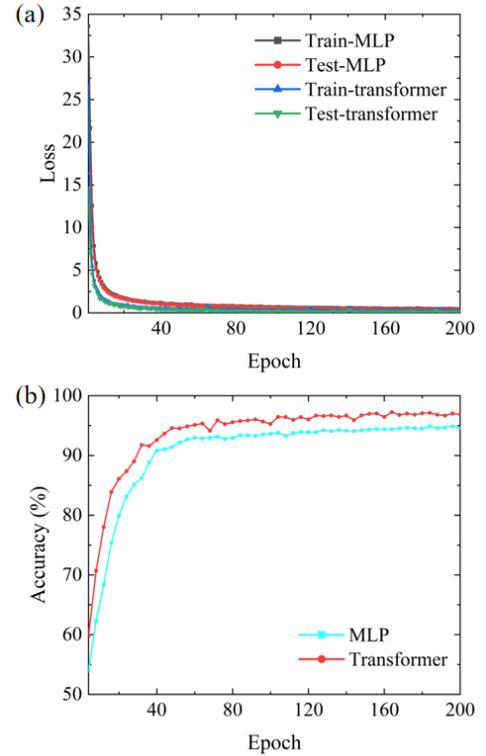

Fig. 4. Performance of StoV design for THz multi-resonant graphene metasurfaces. (a) Training and test loss curves of MLP and improved Transformer. (b) Prediction accuracy of MLP and improved Transformer.

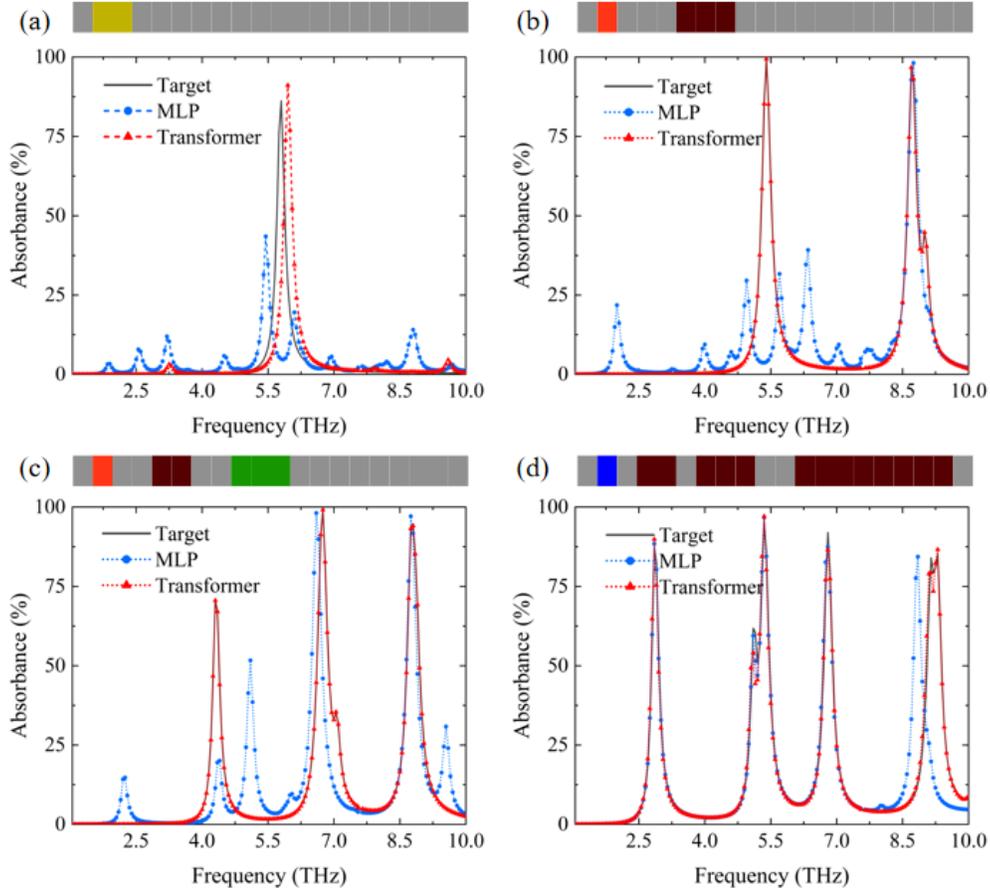

Fig. 5. Absorption spectra of target and predicted based on MLP and improved Transformer algorithms. (a), (b), (c) and (d) represent datasets with one, two, three and four absorption peaks, respectively.

## IV. StoI Design for Graphene Metasurfaces

The StoV design is based on the chemical potential combination of 20 graphene strips, however, the inverse design based on this method is not intuitive enough and cannot achieve inverse design of complex graphene metasurface shapes. Therefore, we further proposed an improved Conditional Generative Adversarial Network (CGAN) to achieve the inverse design of graphene metasurface images directly from the desired multi-resonant absorption spectra, namely StoI design.

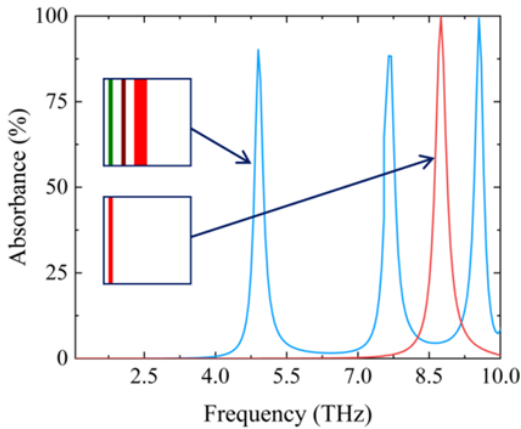

Fig. 6. Typical StoI data with metasurface images and the corresponding absorption spectra.

In the StoI dataset, each metasurface image corresponds to a THz multi-resonant absorption spectrum. The size of the metasurface is 80×80 and the dimension of the absorption spectrum is 181. Two sample metasurface images on the left side of Fig. 6 correspond to blue and red multi-resonant absorption spectra, respectively. In these images, green, yellow, red, blue, and dark red colors represent the five chemical potentials of graphene, namely 0.6 eV, 0.7 eV, 0.8 eV, 0.9 eV, and 1.0 eV, respectively, while white indicates the absence of graphene. This dataset can be extended to include other complex-shaped graphene metasurfaces as long as a sufficient dataset is available for those shapes.

In the original CGAN neural network [33], not only conditional information is required as input, but also random noise signals, which can lead to poor learning of the THz absorption spectrum, thereby affecting the accuracy of the inverse design. However, the improved CGAN directly inputs the THz absorption spectrum to the generator as conditional information to guide the model in generating metasurface images with specific characteristics. The structure of the improved CGAN neural network is shown in Fig. 7, which is a type of generated adversarial neural (GAN) network with conditional constraints. In the improved CGAN, the THz multi-resonant absorption spectrum is used as a condition to guide the generation process of graphene metasurfaces in the model. In this way, the generator can better learn the THz absorption spectrum and generate more accurate metasurface. During the training process, the generator would generate the metasurface samples by THz absorption spectrum, while the discriminator would judge the realism of the generated metasurfaces by comparing them with the real metasurfaces. Through continuous iterative training, the generator would gradually learn how to generate more realistic metasurface images, and the discriminator would become more accurate in judging the generated metasurfaces and the real metasurfaces, thus improving the quality of the overall model generation.

In the modified CGAN, the specific layers of the generator network and the discriminator network are illustrated in Table I. The generator network consists of a fully connected layer and five transposed convolutional layers, and the discriminator consists of two fully connected layers and three convolutional layers. In the generator network, the THz multi-resonant absorption spectrum with 181 sampling points is first passed through a fully connected layer, expanding the input to 5×5×128. Then, this is converted into an image of size 5×5 with 128 channels. Next, the image is upsampled to size 80×80 with 32 channels using four transposed convolution layers. In the last transposed convolution layer, the image size is unchanged and the number of channels is reduced to 3, thus achieving the generation of an 80×80 image from a THz multi-resonant absorption spectrum with 181 sampling points. In the discriminator network, the absorption spectrum is first transformed into a long sequence of 3×80×80 by a fully connected layer, and then converted into an image with a size of 80×80 and 3 channels. The generated image by the generator is concatenated with the absorption spectrum image by channel, resulting in an image with a size of 80×80 and 6 channels. Second, the resulting image is passed through three convolutional layers, resulting in an output dimension of 10×10×64. Finally, it is outputted to one dimension through a fully connected layer, which is applied to identify whether the input image is real or fake.

To train and validate the improved CGAN neural network in this work, we choose the Adam optimizer, and the weight decay is set as 1e-5, while the batchsize is set as 128. The learning rate for the generator and discriminator is set to 0.0005 and 0.00005, respectively. When calculating the accuracy of the neural network, the metasurfaces generated by the CGAN were simulated using the FEM method to obtain the corresponding THz absorption spectra. Then, Formula (14) was employed to calculate the difference between the target spectrum and the THz absorption spectrum generated from the inverse design.

Next, we plotted the training loss curve and the prediction accuracy curve of the CGAN network as shown in Fig. 8. The model converged at epoch 17 with a generator loss of 0.68 and a discriminator loss of 1.33, as reflected in Fig. 8(a). Next, Figure 8(b) shows that the CGAN model achieved a test accuracy of 95.34%. In the inverse design of graphene-based THz metasurfaces, the CGAN model demonstrates higher accuracy compared to the StoV inverse design achieved through MLP neural network. It is the main reason that the CGAN is capable of learning more information from the StoI dataset, which leads to higher accuracy in inverse design and allows the model to reach stability more quickly.

In Fig. 9, we present four THz multi-resonant metasurfaces based on graphene to evaluate the performance of CGAN in inverse design for more specific samples. A comparison is depicted between the target spectrum and the predicted spectrum. The two small images on the left show the desired metasurface image (top) and the metasurface image generated by the CGAN neural network (bottom), respectively, while the spectra of the target metasurface and the metasurface generated by the CGAN are shown in the main figure. The results of the inverse design of StoI demonstrate that the modified CGAN neural network exhibits a more intuitive inverse design effect compared to the traditional MLP neural network. It is capable of directly generating the desired graphene metasurface and has a stronger adaptability in inverse designs of complex-shaped graphene metasurfaces.

TABLE I
SPECIFIC LAYERS IN THE GENERATOR AND DISCRIMINATOR STRUCTURE.

| $i_{th}$ Layer | Generator | Discriminator |
| --- | --- | --- |
| 1 | Linear (181, 5*5*128) | Linear (181, 3*80*80) |
| 2 | ConvTranspose2d (128, 256, 4, 2, 1) | Conv2d (6, 16, 3, padding=1) |
| 3 | ConvTranspose2d (256, 128, 4, 2, 1) | Conv2d (16, 32, 3, padding=1) |
| 4 | ConvTranspose2d (128, 64, 4, 2, 1) | Conv2d (32, 64, 3, padding=1) |
| 5 | ConvTranspose2d (64, 32, 4, 2, 1) | Linear (64*10*10, 1) |
| 6 | ConvTranspose2d (32, 3, 3, 1, 1) | |

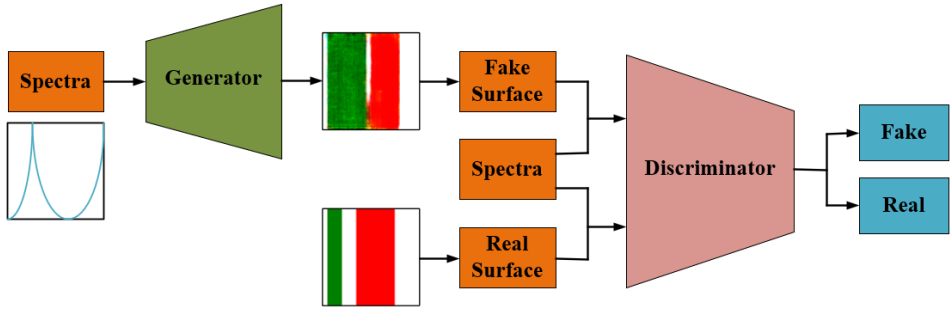

Fig. 7. Structure of the proposed improved CGAN network.

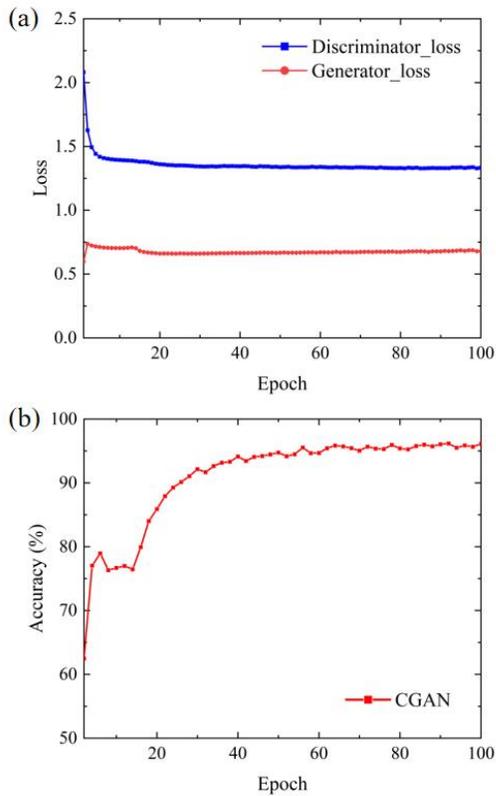

In order to comprehensively compare the performance of three neural networks in implementing inverse design of different types of graphene metasurfaces, we have listed the corresponding inverse design performance in Table II. It can be found intuitively that in the StoV inverse design, the improved Transformer neural network has higher accuracy than the traditional MLP neural network. Besides, the CGAN neural network can learn the two-dimensional graphene metasurface through the StoI dataset, which has higher accuracy in inverse design. Moreover, the model can achieve stability more quickly and is also applicable to more complex inverse design of graphene metasurfaces.

TABLE II
COMPARISON OF THE INVERSE DESIGN OF THREE DIFFERENT NEURAL NETWORKS IN THE StoV AND StoI DATASETS.

| Data Set | Model | Accuracy | Epoch |
| --- | --- | --- | --- |
| StoV | MLP | 94.27 | 33 |
| StoV | Transformer | 96.14 | 16 |
| StoI | CGAN | 95.34 | 17 |

Fig. 8. Performance of StoI design for THz multi-resonant graphene metasurfaces. (a) Training loss curves of discriminator and generator. (b) Prediction accuracy of CGAN.

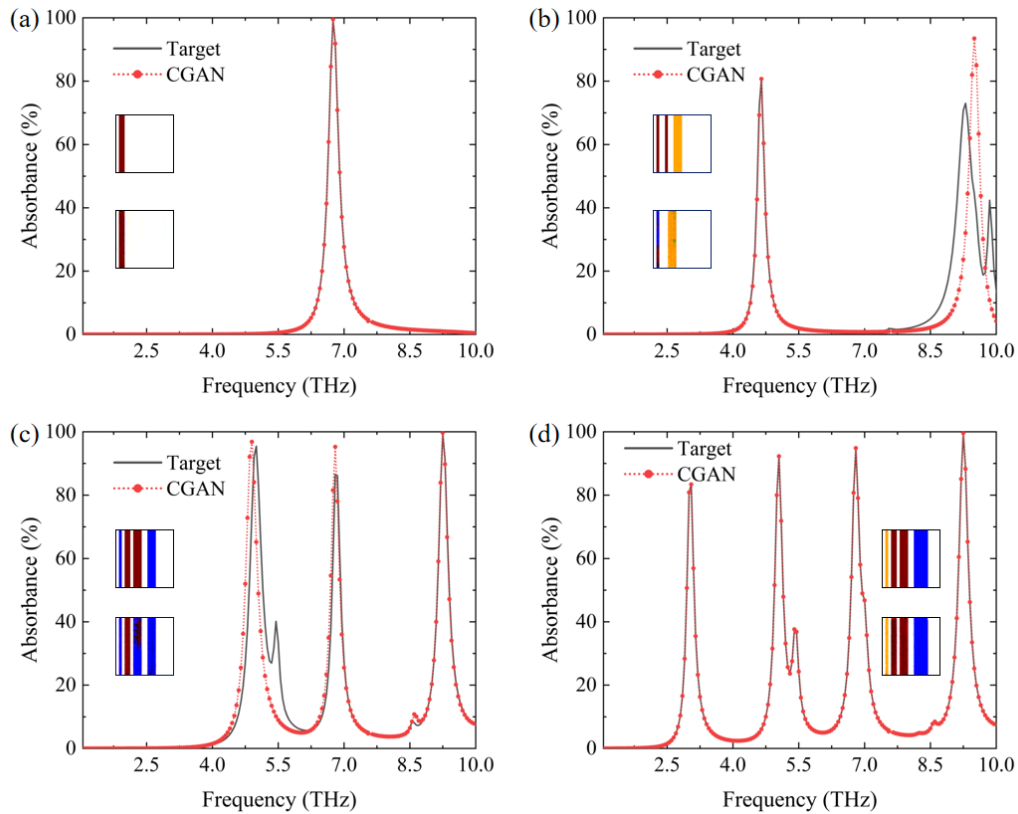

Fig. 9. Metasurface images and their corresponding absorption spectra, both for the target and the predicted values.

## V. CONCLUSION

In this work we mainly investigated two improved neural networks for designing graphene-based THz multi-resonant metasurfaces on-demand. The improved Transformer and CGAN networks were validated by using the StoV and StoI datasets, respectively. It is shown from the results that the improved Transformer network has higher accuracy and faster convergence speed than the traditional MLP neural networks in the StoV design. The improved CGAN achieved higher inverse design accuracy through StoI dataset and provided a more intuitive inverse design process, which can also be employed to design other complex graphene-based metasurfaces in the microwave or millimeter wave range.

## ACKNOWLEDGMENT

The authors declare no conflicts of interest regarding this article.

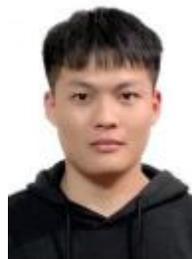

**Yangpeng Huang** is a student of Opto-Electronic and Communication Engineering, Xiamen University of Technology, China. His current research interests are mainly in the areas of graphene metamaterials, nanophononics, inverse design of terahertz metamaterials, deep learning and its applications. He is familiar with various deep learning algorithms and models, such as convolutional neural networks, self-supervised learning, generative adversarial networks, Transformers.

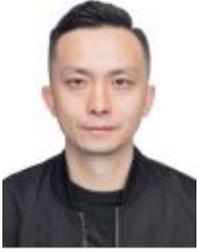

**Yijun Cai** (Member, IEEE) received the B.S. degree in electronic engineering and the Ph.D. degree in physical electronics from Xiamen University, Xiamen, China, in 2010 and 2015, respectively. He currently is an Associate Professor of School of Opto-Electronic and Communication Engineering, Xiamen University of Technology, China. He has authored and coauthored more than 40 peer-reviewed journal and conference papers. His current research interests are mainly in the areas of terahertz metamaterials, nanophotonics, artificial intelligence and its applications.

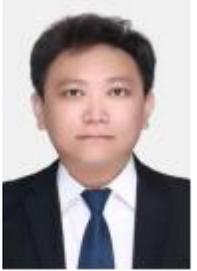

**Naixing Feng** (Senior Member, IEEE) received the B.S. degree in Electronic Science and Technology and the M.S. degree in Micro-Electronics and Solid-State electronics from Tianjin Polytechnic University, Tianjin, China, in 2010 and 2013, respectively. He studied and worked as a Visiting Scholar supported by the CSC in the department of electrical and computation engineering from 2015 to 2016 in Duke University, USA. He received his Ph.D. degree in the College of Electronic Science and Technology from Xiamen University, Xiamen, China, in 2018.

From September 2018 to September 2021, he was an Associate Professor with the Shenzhen University. From July 2020 to June 2022, he was a Visiting Research Fellow with the Changchun Institute of Optics, Fine Mechanics and Physics, Chinese Academy of Science. Since Nov. 2021, he has been an Associate Professor with the School of Electronic Information and Engineering, Anhui University.

He has appeared around 65 papers published by refereed international journals and conference and is the holder of 6 patents. His current research interests contain computational Electromagnetics, machine learning, metamaterials, and van der Waals materials.